\documentclass[sn-mathphys]{sn-jnl}

\jyear{2021}%

\theoremstyle{thmstyleone}%
\theoremstyle{thmstyletwo}%

\theoremstyle{thmstylethree}%

\begin{document}

\title[Partial Ly$\alpha$ thermalization in cosmology]{Partial Ly$\alpha$ thermalization in an analytic nonlinear diffusion model}


\author*{\fnm{Georg} \sur{Wolschin}}\email{wolschin@uni-hd.de}



\affil{\orgdiv{Institute for Theoretical Physics}, \orgname{Heidelberg University}, \orgaddress{\street{Philosophenweg 16}, \city{Heidelberg}, \postcode{69120}, \state{Baden W\"urttemberg}, \country{Germany}}}




\abstract
{During recombination, the cosmic background radiation is disturbed, in particular, by Lyman-alpha emissions from neutral hydrogen. It is proposed to account for the subsequent time-dependent {partial} thermalization of the Ly$\alpha$ energy content in an analytically solvable nonlinear diffusion model. The amplitude of the partially thermalized and redshifted Ly$\alpha$ line is found to be too low to be visible in the cosmic microwave spectrum, in accordance with previous numerical models and Planck observations.}




\maketitle

\section{Introduction}
Following the predictions \cite{reg33,dic64}, and {discovery} of the cosmic microwave background (CMB) with a temperature of $(3.5\pm1.0)$ K at a frequency of $4080$ MHz  \cite{pewi64}, its spectrum has been measured with ever increasing precision. The radiation has a Planck distribution, because the cosmic background radiation (CBR) had been thermalized essentially through Compton scattering and bremsstrahlung  \cite{chan75} at very early times corresponding to redshifts $z>10^7$, and expansion retains the thermal spectrum.  Ground-, balloon-, and rocket-based observations confirmed the low-frequency Rayleigh--Jeans branch of the CMB blackbody distribution. Using the COBE satellite's far-infrared absolute spectrophotometer (FIRAS) \cite{ma90}, it became possible to measure across the peak at $\nu_\text{peak}\simeq 150$ GHz in the frequency range $60$ GHz $\lesssim \nu \lesssim 600$ GHz, and determine the average temperature as $T_\text{CMB}=(2.725\pm0.001)$ K \cite{fixma02}. With COBE's differential microwave radiometer (DMR), statistically significant  {spatial} structure was discovered and described as scale-invariant fluctuations. After subtracting the dipole anisotropy of order $\Delta T / T \simeq 10^{-3}$  that is caused by the motion relative to the CMB, 
primordial temperature fluctuations $\Delta T / T \simeq 6\times10^{-6}$
were found \cite{smo92}, and later measured in great detail with the WMAP \cite{ben13} and Planck \cite{planck20} spacecraft down to very small angular scales of approximately $0.1^\circ$ that correspond to the physical scale of galaxies and clusters of galaxies. 

Although present research emphasizes the implications of the primordial spatial temperature fluctuations for the cosmological models and, in particular, for the determination of the                          $\Lambda$CDM-parameters that are decisive for the fraction of dark matter and dark energy in the universe, it is also of interest to investigate how {frequency} perturbations of the blackbody spectrum disappear, or persist, as function of time \cite{wey66,zs69}. 
Such perturbations occur, in particular, in the course of the recombination epoch of the evolution of the universe, {as investigated and reviewed e.g. in \cite{sc09} with an emphasis on the associated release of photons during this epoch that can cause
deviations of the CMB spectrum from a perfect blackbody, which could in principle be observable in precise measurements. The presently available models such as \cite{zs69,sc09} rely on transport equations that can only be solved numerically.}

In this work, it is proposed to account for the  {incomplete} time-dependent thermalization of the Ly$\alpha$ emissions following recombination through solutions of a nonlinear boson diffusion equation (NBDE) \cite{gw18}. It has proven to be useful in the context of fast gluon thermalization in relativistic heavy-ion collisions \cite{gw22a,gw22} and Bose--Einstein condensate formation in ultracold atoms \cite{gw22a,kgw22}.  {Now the effect of the partially thermalized Ly$\alpha$ line on the CMB spectrum is investigated.} Although many other spectral lines from hydrogen and helium are emitted during recombination, the Ly$\alpha$ line from hydrogen is the most intense.
During recombination, about 68\% of the spectral lines that are emitted by the emerging neutral hydrogen atoms are Lyman-alpha lines arising from $2p_{3/2}\rightarrow 1s_{1/2}$  and $2p_{1/2}\rightarrow 1s_{1/2}$ transitions in neutral hydrogen at frequencies of 2466.071 THz and 2466.060 THz, respectively \cite{md14,kra10}. Like other photons that are emitted following recombination, the ultraviolet  Ly$\alpha$ emissions constitute a disturbance of the high-frequency (Wien) side of the spectrum. Most of the emissions will be re-absorbed and re-emitted by other neutral hydrogen atoms, but {partially thermalize in the course of time through Thomson scattering from the remaining free electrons and other processes. They persist in a nonequilibrium state that could possibly be detected through frequency modulations of the CMB spectrum, or may lead to modified cosmological parameters.}

The nonlinear boson diffusion equation (NBDE) \cite{gw18,gw22a} accounts for time-dependent (partial or complete) thermalization towards the Bose--Einstein stationary distribution that is reached in the limit $t\rightarrow\infty$. 
 In the general case of frequency-dependent transport coefficients, it can only be solved numerically, but analytical solutions exist for constant drift and diffusion coefficients. It is one of the few nonlinear partial differential equations with a clear physical meaning that has analytical solutions. In this work, these solutions are applied to the  {partial} thermalization of the Ly$\alpha$ line that is emitted at recombination, corresponding to a redshift of the last-scattering surface $z_\text{rec}\simeq1100$, and an average recombination temperature of $T_\text{rec}\simeq3000$ K.

The focus is on the implementation of the model into the cosmological scenario using phenomenological values for the drift coefficient $J$ and the {associated} diffusion coefficient $D$. 
As in a more general model with frequency- and time-dependent transport coefficients, these are related to the equilibrium temperature through a fluctuation--dissipation relation $T=-\lim_{t\rightarrow\infty}D(\nu,t)/J(\nu,t)$, thus constraining the value of the drift once the diffusion coefficient has been determined, and vice versa. 
The thermalization timescale is $\tau_\text{eq}\propto D/J^2$, and the proportionality factor will eventually have to be derived from astrophysical input, or could probably be measured in a laboratory. So far, no upper or lower limits are known. In this work, the drift coefficient $J$ will be estimated on phenomenological grounds,
and the diffusion coefficient is computed from the fluctuation-dissipation relation with the equilibrium temperature T.

The nonlinear boson diffusion model is adapted to the cosmological scenario in the next section, and the analytic solution in case of constant transport coefficients is reconsidered. In particular, the initial conditions for the specific case of Ly$\alpha$ thermalization in the early universe 
are incorporated into the analytical solution scheme. In Sect. 3,  the time-dependent results of the thermalization problem for the Ly$\alpha$ initial conditions are presented, and
the relation of the transport coefficients to the equilibration timescale is discussed. As a model calculation that does not yet reflect the physically realistic situation in cosmology, the case of complete thermalization is investigated in Sect. 4, where it is shown that the solutions of the NBDE correctly approach the Bose-Einstein limit for large times. In Sect. 5, the case of incomplete thermalization is discussed that corresponds to the actual time evolution of the Ly$\alpha$ line from recombination in cosmology.  An upper limit for the effect of the partially thermalized Ly$\alpha$ line from recombination on the CMB is calculated. The conclusions are drawn in Sect. 6.

\section{Nonlinear boson diffusion model in cosmology}
\noindent
Planck's equilibrium spectrum for the specific intensity (spectral radiance) as function of frequency $\nu$ at  temperature $T$ is
\begin{align}
L(\nu,T)=\frac{2h\nu^3}{c^2}\frac{1}{\exp\left(\frac{h\nu}{k_\text{B}T}\right)-1}\,.
\label{planck}
\end{align}
Due to the expansion of the universe the equilibrium temperature $T$ decreases, but the thermal spectrum is maintained, because both temperature and frequency are reduced with redshift as $(1+z)$  such that $\nu/T$ is unchanged. The chemical potential $\mu$  may initially be smaller than zero, but is driven towards zero in the course of time 
and hence, does not appear in the thermal spectrum. The cosmic background radiation at early times as well as the CMB radiation at present are therefore modelled as blackbody spectra with $\mu=0$, as in eq.\,(\ref{planck}).
\begin{figure}[H]
\centering
	\includegraphics[width=0.8\columnwidth]{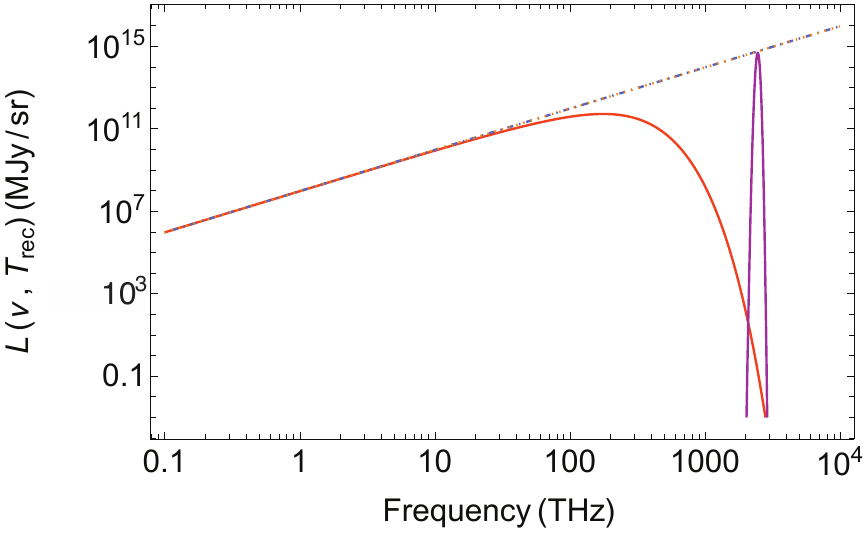}
    \caption{Planck spectrum of blackbody radiation at recombination with an average temperature of $T_\text{rec}=3000$ K at redshift $z_\text{rec}\simeq 1100$, and the thermally broadened Lyman-$\alpha$ line of neutral hydrogen at $\nu_\alpha=2466$ THz in the VUV region of the spectrum. The dot-dashed line is the Rayleigh--Jeans distribution, $L_\text{RJ}(\nu,T)\propto \nu^2T_\text{rec}$.}
    \label{fig1}
\end{figure}

\begin{figure}[H]
\centering
	\includegraphics[width=0.8\columnwidth]{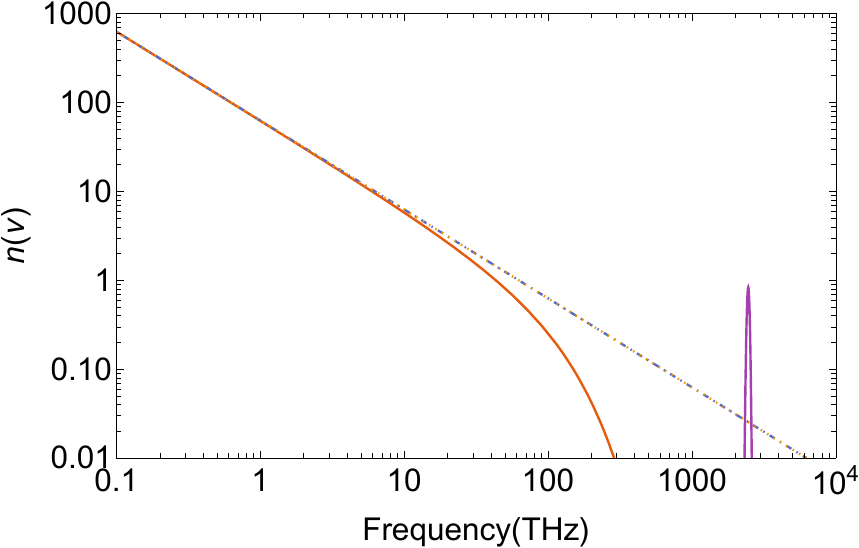}
    \caption{Bose--Einstein occupation-number distribution at recombination with an average temperature of $T_\text{rec}=3000$ K, and the thermally broadened Lyman-$\alpha$ line of neutral hydrogen at $\nu_\alpha=2466$ THz, here with normalization $N_\alpha=100$ THz for better visibility. The dot-dashed line is the Rayleigh--Jeans distribution.}
    \label{fig2}
\end{figure}
At the time of recombination  {$\tau_\text{rec}$}$\simeq 380$ ky, the CBR intensity
for an average recombination temperature of $T=T_\text{rec}\simeq3000$ K is shown in Fig.\ref{fig1} 
 together with the Lyman-$\alpha$ line at $\nu_\alpha=2466$ THz in the vacuum ultraviolet (VUV) region of the electromagnetic spectrum. This line is a doublet with transition frequencies
$2466.071$ THz and $2466.060$ THz \cite{kra10} corresponding to the respective $2p_{3/2}\rightarrow 1s_{1/2}$  and $2p_{1/2}\rightarrow 1s_{1/2}$ transitions in neutral hydrogen having Lorentzian line profiles. At the recombination temperature, however, 
the natural line widths are thermally broadened, such that a single Gaussian profile of line width $\Gamma_\alpha\simeq1.8\, T_\text{rec}\simeq62.5$ THz can be used to represent the occupation-number distribution of the Lyman-alpha line  
\begin{align}
n_\alpha(\nu)=\frac{N_\alpha}{\sqrt{2\pi}\sigma_\alpha}\exp\left[-\frac{(\nu-\nu_\alpha)^2}{2\sigma_\alpha^2}\right]
\label{ly}
\end{align}
with normalization $N_\alpha$ in THz, and standard deviation $\sigma_\alpha=\Gamma_\alpha/\sqrt{8\ln2}=1.8\,T_\text{rec}/\sqrt{8\ln2}\simeq26.8$ THz. The specific line intensity at recombination $L_\alpha(\nu,T_\text{rec})=2h\nu^3/c^2\times n_\alpha(\nu,T_\text{rec})$ normalized to Rayleigh--Jeans is shown in Fig.\ref{fig1}. A physically realistic normalization will eventually have to rely on observational data, but in Sec.\, 5, an upper-limit estimate will be given.
 In the following, a system of units $h/(2\pi)\equiv\hbar=c=k_\text{B}=1$ will be used.
 
The position of a redshifted Ly$\alpha$ line in the CMB spectrum would be at about $2240$ GHz, in the far Wien end. It is, however, likely that the line is partially thermalized until today through random scatterings with the remaining free electrons (few parts in $10^4$ after recombination), {resonance scattering,}
and other processes.
If the timescale for thermalization turns out to be comparable to, or larger than the inverse expansion rate, the expansion during thermalization has to be taken into account. {Since} Ly$\alpha$ thermalization is {likely} not completed until the present time, a far-infrared nonthermal background (FIRB) could survive in the CMB above several hundred GHz. However, other nonthermal sources may also contribute there, such as light generated by star formation that is absorbed by interstellar dust and re-emitted in the far infrared. Indeed, FIRB radiation had been detected by COBE's FIRAS \cite{cobe96} and DIRBE \cite{cobe98} instruments, but it is difficult to disentangle its various possible sources.

Once the diffusion function $D(\nu,t)$ and the drift function $J(\nu,t)$ that account for 
thermalization are known, the time-dependence of the photonic single-particle occupation number distribution $n(\nu,t)$ from its initial distribution eq.(\ref{ly}) towards the equilibrium distribution $n_\infty(\nu)$ can be calculated from solutions of the nonlinear boson diffusion equation \cite{gw18,gw22a} 
 \begin{align}
 \label{nbde}
		\frac{\partial n}{\partial t}=-\frac{\partial}{\partial{\nu}}\left[J\,n\left(1+ n\right)+n\frac{\partial D}{\partial \nu}\right]
	+\frac{\partial^2}{\partial{\nu}^2}\bigl [D\,n\,\bigr]\,.
\end{align}
The mean occupation number $n(\nu,t)$ is equal to the mean energy, divided by the energy per photon.
This nonlinear diffusion equation has been derived for bosonic systems in \cite{gw18} to account for the fast thermalization of gluons in relativistic heavy-ion collisions, but it also accounts for thermalization in other nonequilibrium Bose systems, such as cold atoms, or photons. To obtain the NBDE, the quantum Boltzmann equation is first rewritten in form of a master equation,  and the discrete transition probabilities between quantum states are expressed as integrals by introducing the corresponding densities of states. An approximation to the master equation is then obtained through a gradient expansion in energy space, and drift and diffusion coefficients are introduced as first and second moments of the transition probabilities, respectively, to finally arrive at the above nonlinear 
eq.\,(\ref{nbde}).

The drift term $J(\nu,t)$ in the NBDE is negative. It is mainly responsible for dissipative effects {such as recoil \cite{gd08}} that drive the distribution towards lower frequencies and cause boson (photon) enhancement, the diffusion term $D(\nu,t)$ accounts  {via the fluctuation-dissipation theorem} for the diffusion in frequency (energy) space.
The derivative-term of the diffusion coefficient is required 
such that the stationary solution $n_\infty(\nu)$ becomes a Bose--Einstein equilibrium distribution.
This can be seen by rewriting the equation, setting the time derivative
to zero and solving for $n_\infty$.

With the condition that the ratio $J(\nu,t)/D(\nu,t)$ must have no frequency (energy) dependence for ${t\rightarrow\infty}$ such that
$\lim \limits_{t \to \infty}[-J(\nu,t)/D(\nu,t)] \equiv 1/T$,
it can be shown \cite{gw22}
that the stationary distribution $n_\infty$ equals the Bose--Einstein equilibrium distribution $n_\text{eq}$, respectively, 
\begin{align}
n_\infty(\nu)=n_\text{eq}(\nu)=\frac{1}{e^{(2\pi\nu-\mu)/T}- 1}\,.
 \label{Bose--Einstein}
\end{align}
Here, the chemical potential $\mu\le 0$ appears as a parameter. In Fig.\,\ref{fig2}, the equilibrium distribution for $\mu=0$ is shown at recombination together with the Rayleigh-Jeans distribution and the thermally broadened Ly$\alpha$ line.

  
The nonlinear diffusion equation for the occupation-number distribution $n(\nu,t)$ becomes particularly simple for frequency-independent transport coefficients
\begin{align}
	\frac{\partial n}{\partial t}=-J\,\frac{\partial}{\partial{\nu}}\Bigl[n\,(1+ n)\Bigr]+D\,\frac{\partial^2n}{\partial{\nu}^2}\,,
	\label{bose}
\end{align}
where the derivative-term of the diffusion coefficient is now absent, and the transport coefficients have been pulled in front of the derivatives.

The equation with constant transport coefficients differs significantly from a linear Fokker--Planck equation -- which has the Maxwell--Boltzmann distribution as stationary solution -- due to the nonlinear term: It 
preserves the essential features of Bose--Einstein statistics that are contained in the quantum Boltzmann equation. This refers especially to the Bose enhancement in the low-frequency region that increases rapidly with time. Indeed, for ultracold bosonic atoms it has been shown in \cite{gw22a,kgw22} that the simplified equation with constant transport coefficients -- together with the requirement of particle-number conservation -- already accounts for time-dependent condensate formation in agreement with recent Cambridge data  \cite{gli21}. At much higher energies and temperatures, the NBDE has been used in \cite{gw22a,gw22} to account for the fast thermalization of gluons in relativistic heavy-ion collisions at energies reached at CERN's large hadron collider (LHC). 

The diffusion equation with constant coefficients can be solved in closed form for any given initial condition $n_\mathrm{0}(\nu)$  using the nonlinear transformation outlined in \cite{gw18,gw22a}. 
 The resulting exact solution of the NBDE can be expressed as
\begin{align}
        n\,(\nu,t) = T\, \partial_\nu\ln{\mathcal{Z}(\nu,t)} - \frac{1}{2}= \frac{T}{\mathcal{Z}}\, \partial_\nu \mathcal{Z} -
         \frac{1}{2}\,,
    \label{net} 
    \end{align}
 where the generalized (time-dependent) partition function ${\mathcal{Z}(\nu,t)}$ obeys a linear diffusion (heat) equation 
  \begin{align}
  \frac{\partial}{\partial t}{\mathcal{Z}}(\nu,t) = D \frac{\partial^2}{\partial \nu^2}{\mathcal{Z}}(\nu,t)\,.
  \label{eq:diffusionequation}
  \end{align}
The time-dependent
partition function can be written as an integral over Green's function of the above eq.\,(\ref{eq:diffusionequation}) and an exponential function $F(x)$ 
which depends on
 the initial occupation-number distribution $n_\mathrm{0}$
according to
 \begin{align}
    F(x) = \exp\Bigl[ -\frac{1}{2D}\bigl(J x+ 2J \int^x n_\text{0}(y)\,{d}y \bigr) \Bigr]\,.
       \label{fini}
\end{align}
Here, the integration constant can be omitted, because it will drop out once the logarithmic derivative of the partition function is taken.
The time-dependent partition
function 
with boundary conditions at the singularity becomes
     \begin{align}
    \mathcal{Z}_\mathrm{bound} (\nu,t)= \int_0^{+\infty} G_\mathrm{bound} (\nu,x,t)\,F(x)\,{d}x\,.
    \label{eq:partitionfunctionZ}
    \end{align}

For sufficiently simple initial conditions, it can be calculated  analytically, as has been done in \cite{rgw20} for ultracold atoms. 
In eq.\,(\ref{eq:partitionfunctionZ}), Green's function $G_\mathrm{bound}$ accounts for the boundary conditions at the singularity $2\pi\nu=\mu=0$. They can be expressed as
 $\lim_{\nu \downarrow 0} n\,(\nu,t) = \infty~ \forall ~t$. One obtains a vanishing partition function -- corresponding to an infinite occupation-number distribution -- at the boundary,
  \( \mathcal{Z}_\text{bound} (\nu=0,t) = 0\), and the energy range is restricted to  $\nu\ge 0$. This requires a Green's function 
that equals zero at $\nu = 0~ \forall \,t$. It can be written as
\begin{align}
    {G}_\mathrm{bound} (\nu,x,t) = G_\mathrm{free}(\nu,x,t) - G_\mathrm{free}(\nu,-x,t)\,,
    \label{eq:newGreens}
\end{align}
with the free Green's function $G\equiv G_\mathrm{free}$ of the linear diffusion eq.\,(\ref{eq:diffusionequation}),
\begin{align}
		G_\mathrm{free}(\nu,x,t)=\frac{1}{\sqrt{4\pi Dt}}\exp{\left[- \frac{(\nu-x)^2}{4Dt}\right]}\,.
	\label{eq:Greensnonfixed}
\end{align}
 Finally, the occupation-number distribution is obtained via the basic nonlinear transformation, eq.\,(\ref{net}).
\section{Time-dependent calculations}
To account for the time-dependent {-- partial or complete --} thermalization of the Ly$\alpha$ energy content subsequent to recombination, the integral over the initial distribution of the Ly$\alpha$ line is obtained as
 \begin{align}
  \int^x n_\text{0}(y)\,{d}y\,&=  \int^x \frac{N_\alpha}{\sqrt{2\pi}\sigma_\alpha}\exp{\left[\frac{(y-\nu_\alpha)^2}{2\sigma_\alpha^2}\right]}dy       \label{ini}\\ \nonumber
& = \frac{N_\alpha}{2\sigma_\alpha}\text{erf}\left[\frac{x-\nu_\alpha}{\sqrt{2}\sigma_\alpha}\right]\,,
\end{align}
which is inserted into the exponential function $F(x)$ in eq.\,(\ref{fini}). With Green's function from eq.\,(\ref{eq:newGreens}), the partition function is obtained from eq.\,(\ref{eq:partitionfunctionZ}), and the time-dependent occupation-number distributions can be computed  from the nonlinear transformation, eq.\,(\ref{net}). 

For a realistic calculation of the time-dependent thermalization, the values of the transport coefficients are decisive. They are related to the equilibrium temperature through a fluctuation--dissipation relation $T=-D/J$.
 Moreover, the equilibration time scale is related to the transport coefficients \cite{gw22a} according to
$\tau_\text{eq}=a_\tau D/J^2$ with a proportionality constant $a_\tau$. Hence, the transport coefficients are obtained as
 \begin{align}
D&=a_\tau\,T^2/\tau_\text{eq} \,,\label{diff}\\
J&=-a_\tau\,T/\tau_\text{eq}\label{dri}\,.
\end{align}
{The proportionality factor $a_\tau$ for thermalization in a Bose system depends significantly on the initial condition. So far, it has been calculated analytically only for a $\theta$-function initial distribution that overlaps with the low-frequency branch of the thermal distribution \cite{gw18}. No derivation is available for the present case, where a narrow initial distribution at the UV side of the spectrum is far away from the thermal Bose enhancement at the opposite side of the spectrum, and hence, $a_\tau$ and                       $\tau_\text{eq}$ are treated as parameters that have to be determined in the cosmological context. Alternatively, the drift coefficient is taken as a parameter -- see the next section --, and the diffusion coefficient is computed from the fluctuation-dissipation relation at temperature $T$.} 

With the above connections between the transport coefficients $D, J$, the equilibrium temperature  $T$, and the equilibration time scale $\tau_\text{eq}$, the nonlinear diffusion eq.\,(\ref{bose}) in the limit of constant coefficients can also be expressed using the dimensionless time-like variable $\delta$ 
\begin{align}
		\frac{\partial n}{\partial \delta}=Ta_\tau\frac{\partial}{\partial{\nu}}\Bigl[n\,(1+ n)\Bigr]+{T^2}a_\tau\frac{\partial^2n}{\partial{\nu}^2}\,.
	\label{bose1}
\end{align}
Here, the time variable $t$ of eq.\,(\ref{bose}) has been replaced, $t\rightarrow t_\text{0}+\tau_\text{eq}\delta$. In this dimensionless form of the nonlinear diffusion equation, the temperature $T$ and the equilibration timescale appear as parameters instead of the transport coefficients $J$ and $D$, together with the dimensionless constant $a_\tau$ that characterises thermalization in a Bose system for a given initial condition. {The solution method outlined above for the NBDE with constant transport coefficients can be applied to eq.\,(\ref{bose1}) as well}.
\section{Complete thermalization}
 {The analytic solutions of the nonlinear diffusion equation are first applied to an idealized situation of complete thermalization of the Ly$\alpha$ line during the time evolution. This is, of course, not realistic when accounting for the physics of recombination and the subsequent partial frequency redistribution \cite{sc09} of the photons that are emitted following hydrogen and helium recombination: Scattering is known to hardly be able to thermalize the Ly-alpha distortion even at earlier times. Nevertheless, this calculation serves to demonstrate the method, and it will subsequently be adapted to the actual physical situation of incomplete thermalization of the Ly$\alpha$-line in the next section.}
 
 {The value of the drift coefficient in this schematic calculation is chosen as $J=-1$ THz/ky. The frequency shift of the initial Ly$\alpha$ line at short time intervals $\Delta t \ll\tau_\text{eq}$ is approximately $\Delta \nu_\alpha\simeq J \Delta t$ THz. It later becomes a nonlinear function of time, especially when Bose enhancement sets in at smaller frequencies. Physically, the drift is a consequence of several effects that cause a frequency redistribution towards lower energies such as atomic recoil \cite{gd08} and electron scattering during recombination. Neglecting expansion and cooling for the moment (it will be discussed in the next section), the equilibrium temperature is kept at $T=T_\text{rec}$, and due to the fluctuation-dissipation relation  $D=-T/J$ the corresponding diffusion coefficient becomes $D\simeq 63$ THz$^2$/ky.  }

\begin{figure}[H]
\centering
	\includegraphics[width=0.8\columnwidth]{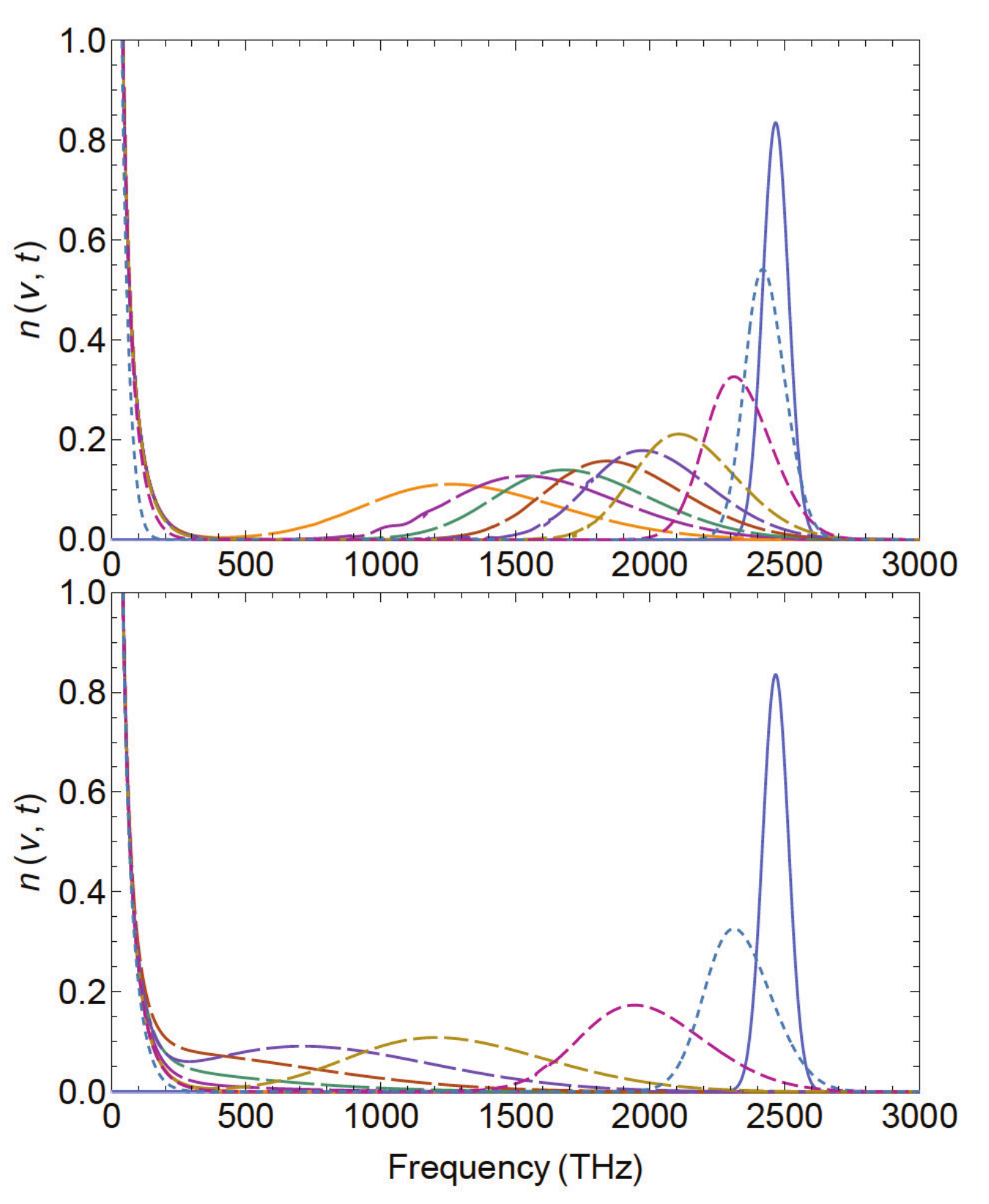}
    \caption{Time-dependent thermalization of the Ly$\alpha$ line (blue, right at 2466 THz) towards the Bose--Einstein occupation-number distribution (solid, left) as calculated from solutions of the nonlinear boson diffusion equation (NBDE). Results are shown for $t/10^3$ ky $=0.1; 0.4; 1.1; 1.5; 2; 2.5; 3; 4$ (increasing dash lengths). {The lower frame is for complete thermalization with $D=63$ THz$^2$/ky and $J=-1$ THz/ky, in the upper frame both transport coefficients are reduced by a factor of 4, causing slower thermalization.}}
    \label{fig3}
\end{figure}

\begin{figure}[H]
\centering
		\includegraphics[width=0.8\columnwidth]{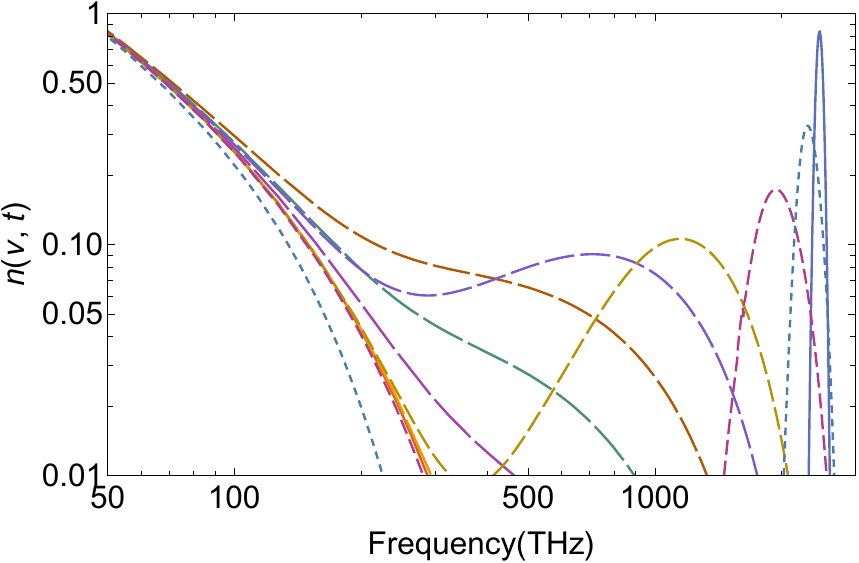}
    \caption{Time-dependent evolution of the Ly$\alpha$ line in the ultraviolet part of the spectrum (blue, right at 2466 THz) towards equilibrium (solid, left) {assuming complete thermalization towards $T=T_\text{rec}$ and time-independent $D, J.$} Eight timesteps are shown as in Fig.\,\ref{fig3} (increasing dash lengths), but in a double-log plot to  {illustrate} the approach towards the thermal occupation-number distribution in the visible and infrared region of the spectrum.}
    \label{fig4}
\end{figure}

{Results of the time-dependent thermalization with the above parameters are shown for eight timesteps in Fig.\,\ref{fig3}, lower frame. The 
Ly$\alpha$ emissions are first broadened and shifted to the low-frequency region, until Bose's enhancement sets in. Equilibrium is reached at 
$t=\tau_\text{eq}\simeq4\times10^3$ ky in this particular example. Dividing both transport  coefficients by a factor of 4 (upper frame) retains the same equilibrium temperature, but thermalization occurs more slowly: It would take $4\times\tau_\text{eq}$ to reach equilibrium, because $\tau_\text{eq}\propto D/J^2$.  
In case of complete thermalization, no remainder of the Ly$\alpha$ emissions from recombination would survive in today's CMB spectrum.
With the consideration of expansion and cooling, and more realistic values for the transport parameters from astrophysical arguments as will be discussed below, the time dependence will differ, and the occupation-number distribution will remain far from equilibrium,} but the principal effects of the approach to equilibrium shall persevere.
The result  {of the lower frame in Fig.\,\ref{fig3}} is shown again for eight timesteps in Fig.\,\ref{fig4} in a double-log plot, which emphasizes the approach to Bose--Einstein equilibrium in the near{-}infrared region. It can also be seen that even at short times, the solutions of the NBDE generate a low-frequency branch (short-dashed curves on the left) that thermalizes very quickly. It is due to the Bose enhancement that is contained in the NBDE, {and leads to a Rayleigh-Jeans-like slope in the infrared when calculating the specific intensity, see the next section}. Beyond the equilibration time scale, the full distribution function of the specific intensity becomes again a Planck spectrum, eq.\,(\ref{planck}).
\section{Incomplete thermalization}

{Whereas the above results show the general viability of the nonlinear diffusion model to account for thermalization in a bosonic (here: photonic) system, the actual Ly$\alpha$ physics in the course of recombination remains far from equilibrium. One important reason is the rapidly falling free-electron number density during recombination, thus diminishing scattering processes that are required for thermalization: At the end of the recombination era at redshift $z\simeq500$ corresponding to an equilibrium temperature of $T{=T_\text{f}}\simeq0.45\,T_\text{rec}$, the relative free-electron density has dropped below $10^{-3}$ \cite{sc09}, the values of the transport coefficients in the NBDE must be reduced accordingly, and their time dependence during recombination has to be considered.}

{The partial frequency redistribution of Ly$\alpha$ photons has been treated in \cite{cs09} based on a Fokker-Planck approximation -- which may, however, not be sufficient towards the end of hydrogen recombination. That numerical approach had been proposed by Rybicki \cite{ry06}, who had also discussed a correspondence to Kompaneets' equation \cite{kom57}  when written in terms of the photon occupation number. This equation concentrates, in particular, on the role of the Compton effect in the establishment of equilibrium between quanta and electrons in a nonrelativistic approximation.} 

{As a complement, the nonlinear diffusion equation offers a related solution to the problem of partial thermalization that properly accounts for Bose statistics, takes into account the boundary conditions, and provides a transparent analytical solution through a suitable nonlinear transformation.}

{First-principles calculations of drift and diffusion coefficients in the NBDE based on the relevant physical processes electron scattering, and resonance scattering off moving atoms -- both including recoil, Doppler broadening and induced scatterings --  are not yet available in the cosmological context for the proposed NBDE-model. The transport coefficients are instead estimated here on phenomenological grounds. {We start from the equilibrium temperature at the 
beginning of the recombination era, $T_\text{i}=T_\text{rec}\simeq3000$ K at redshift $z\simeq1100$. The end of recombination is taken to be reached at $T_\text{f}\simeq 0.45\,T_\text{rec}\simeq 1350$ K at redshift $z\simeq 500$. Using the Planck cosmological parameters $H_0= (67.4\pm0.5)$ km/s/Mpc, $\Omega_\text{m}=0.315\pm 0.007$ \cite{planck20}, these values  correspond to time scales \cite{wri06}
in the $\Lambda$CDM model. In the redshift region $1100\gtrsim z\gtrsim 500$ where partial thermalization is acting in the schematic nonlinear model, the relation between redshift and time can approximately be expressed as an exponential function.}
 
 {Since the temperature depends linearly on the redshift, $T(z)=T_\text{CMB}(1+z)$, a related redshift- or time dependence
is taken for the transport coefficients. For an equilibrium temperature at the end of recombination $T_\text{f}\simeq 1350$ K at redshift $z\simeq500$ and an equilibration time scale $\tau_\text{eq}\simeq 1.5\times 10^3$ ky, this corresponds to exponential time dependencies,
 \begin{align}
D(t)&=D_0\exp(-t/\tau_\text{eq}) \,,\label{DT}\\
J(t)&=J_0\exp(-t/\tau_\text{eq}) \label{JT}\,,
\end{align}
with $J_0=-1$\,THz/ky as in the previous model calculation, and $D_0=-0.45\,T_\text{rec}/J_0=28$ THz$^2$/ky. The ratio $-D(t)/J(t)=T_\text{f}$ remains time independent, as required. The value of the drift may turn out to be somewhat overestimated when compared with detailed numerical simulations of the frequency redistribution such as in \cite{cs09}, but could be adapted accordingly. The value of $D(t)$ is, however, }computed from the fluctuation-dissipation relation, which is inherent to the present model. It could only be modified by permitting {frequency}-dependent transport coefficients, as in eq.\,(\ref{nbde}).

{The result of the time-dependent calculation with the above parameter set is shown in Fig.\,\ref{fig5} in four timesteps up to $t=1.5\times 10^3$ ky, corresponding to redshift $z\simeq500$. The distribution functions become already asymmetric (note the log-scale), indicating that a linear Fokker-Planck approximation may not be justified. They remain, however, far from equilibrium -- except for the low-frequency branch, which thermalizes quickly even for short times, and leads to a Rayleigh-Jeans slope in the specific intensity of the partially thermalized Ly$\alpha$ line, see below.} {Increasing the equilibration time $\tau_\text{eq}$ in eqs.\,({\ref{DT}),(\ref{JT}) artificially by a factor of two causes larger values for drift and diffusion at any given time and thus, somewhat faster partial thermalization (the distribution for $t=1.5\times10^3$ ky peaks at $1290$ THz rather than $1700$ THz), thus decreasing the probability that a remainder of the Ly$\alpha$-line could be visible in the CMB.} }

\begin{figure}[H]
\centering
	\includegraphics[width=0.8\columnwidth]{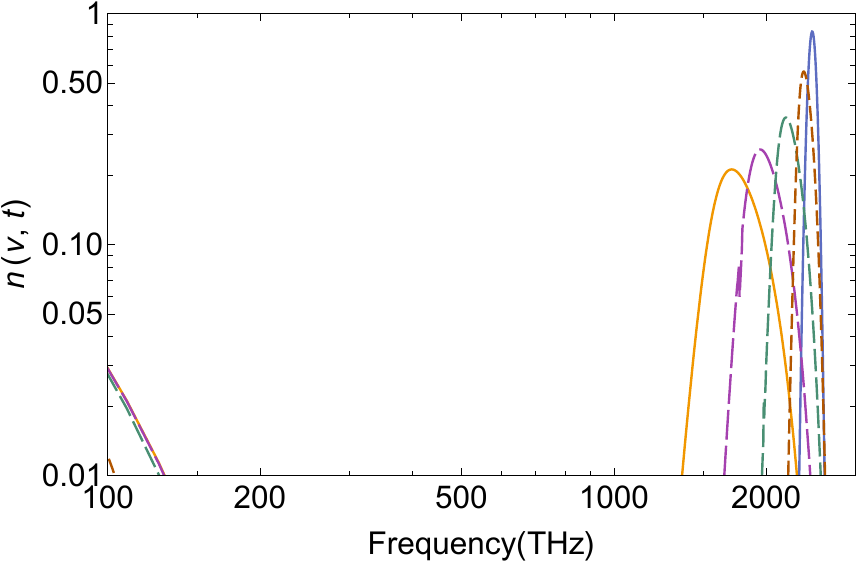}
    \caption{{Nonequilibrium evolution of the Ly$\alpha$ line (blue, right at 2466 THz): Incomplete thermalization with time-dependent transport coefficients $D(t), J(t)$, see text. Timesteps $t/10^3$ ky$=0.05; 0.2;  0.5; 1.5$ are shown, with no significant thermalization expected after the last step, which corresponds to redshift {$z\simeq500$}. The curve at the lower left is the equilibrium distribution for $T=0.45\, T_\text{rec}\simeq 1350$ K, which is reached only at low frequencies.}}
    \label{fig5}
\end{figure}


{Assuming that thermalization indeed terminates at the end of the recombination epoch, the effect of the Ly$\alpha$ photons from hydrogen recombination -- which  represent most of the photons from both, hydrogen and helium ($\simeq24\%$) recombination -- on the CMB distribution based on the NBDE evolution can be estimated by propagating the distribution function at $z=500$ taken from Fig.\,\ref{fig5} (solid yellow curve) to $z=0$. The normalization is taken according to the ratio of photons in the isotropic blackbody radiation to baryons -- mostly protons --, which is approximately $1.6\times10^9$. This yields an upper limit to the effect of the partially thermalized hydrogen Ly$\alpha$  line on distorting today's CMB spectrum.} 

{The result can be seen  in Fig.\,\ref{fig6}, where the blackbody CMB (solid curve) including the COBE-FIRAS data \cite{fixma02} is shown together with the renormalized solution of the NBDE at $z=500$ from Fig.\,\ref{fig5}, propagated to $z=0$ (dashed curve). Its peak resides on the Wien side of the CMB, and the amplitude is more than seven orders of magnitude below the one of the CMB. To test the sensitivity of the nonlinear model, I have reduced the amplitudes $J_0$ of the drift coefficient and $D_0$ of the diffusion coefficient by factors of two, thus keeping the equilibrium temperature at the same value as before. The result is shown as a dotted curve in Fig.\,\ref{fig6}, which is still below the CMB signal by almost seven orders of magnitude.
\begin{figure}[H]
\centering
	\includegraphics[width=0.8\columnwidth]{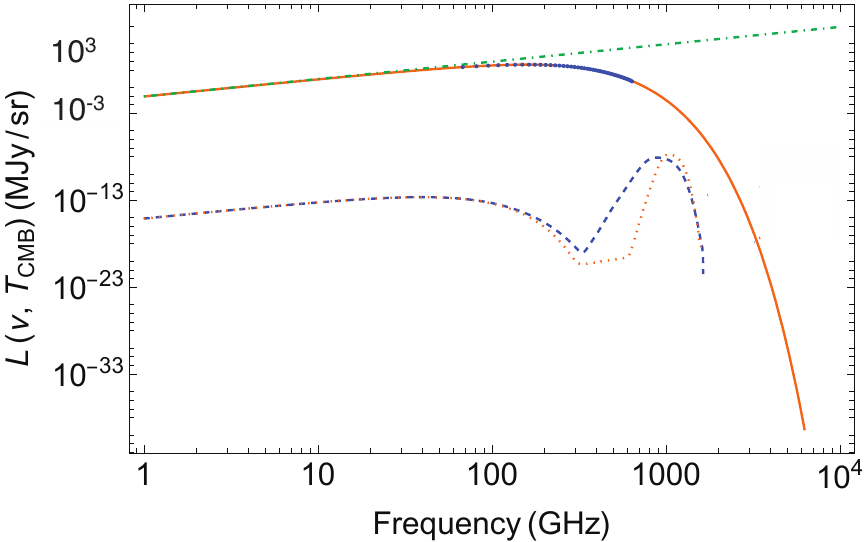}
    \caption{Today's CMB spectrum of blackbody radiation with T=2.725 K (solid curve), the Rayleigh-Jeans distribution (green dot-dashed line), and the partially thermalized NBDE-solution for the redistributed, and redshifted Ly$\alpha$ line featuring a low-frequency R-J branch due to Bose enhancement (dashed curve). The dotted curve is the NBDE-result for a 50\% reduction in the drift and diffusion coefficients. See text for the cutoff at $\nu\simeq2000$ GHz. COBE-FIRAS data \cite{fixma02} are shown as blue dots, error bars are smaller than the symbol size.}
    \label{fig6}
\end{figure}
This largely analytical calculation gives a clear hint why no Ly$\alpha$ signal from recombination -- or oscillatory signal when taking into account all other radiative transitions in hydrogen and helium -- is visible at the
present level of precision in the CMB, although the distortions can lead to biases to several cosmological parameters  \cite{planck20}. }

{As shown in Fig.\,\ref{fig6}, the NBDE result yields a Rayleigh-Jeans slope in the intensity of the Ly$\alpha$ line at low frequencies. This is a consequence of the Bose enhancement, which would not occur in a linear Fokker-Planck type approach to partial thermalization. To reach the proper Wien limit also at large frequencies $\nu>2000$ GHz, energy-dependent transport coefficients would be required -- which is beyond the scope of an analytical model.} 

\section{Conclusions}

The time-dependent {incomplete} thermalization of the Ly$\alpha$ line that is emitted from neutral hydrogen atoms during recombination has been accounted for in a nonlinear diffusion model. {This approach is complementary to the available detailed numerical treatments of the release of photons during the recombination epoch, their partial frequency redistribution, its effect on the recombination history, and possible observable distortions of the CMB. }

In the analytical model, the thermally broadened Ly$\alpha$ emission line at an average recombination temperature of 3000 K {provides the} initial condition. With the proper boundary condition that causes the low-frequency Bose enhancement, the diffusion equation is solved through a nonlinear transformation in the limit of constant transport coefficients. The stationary solution is equal to the Bose--Einstein equilibrium distribution. 

{As is well known, the system remains far from equilibrium, because the interaction of the radiation field with the electrons can not transform a non-Planckian spectrum into a Planckian one in the course of, or after, recombination. 
However, the low-frequency Rayleigh-Jeans slope in the specific intensity indicative for thermalization correctly emerges already at very short times from the analytical solutions of the nonlinear diffusion equation -- which would not be the case in a linear theory for the frequency redistribution of Ly$\alpha$, or other recombination lines. {Variations of the transport coefficients by a factor of two have shown that the model gives robust results. Moreover, it is not overly sensitive to modifications in the time-dependence of the transport coefficients, thus enhancing its reliability in the astrophysical and cosmological context.}

{In the present work, I have investigated partial Ly$\alpha$ thermalization in a redshift range $1100\gtrsim z\gtrsim 500$}. Additional thermalization may result at later times, in particular, during the epoch of reionisation, when the ultraviolet light from the first stars at $t\simeq370$ My and redshift $z\simeq12$ re-ionizes hydrogen and helium until about $z\simeq6$. {This may be another interesting topic for the application of the nonlinear diffusion model. }

Regarding possible signatures of the redshifted Ly$\alpha$ recombination line in today's CMB, an upper limit of the specific intensity following partial frequency redistribution as calculated from the nonlinear diffusion model is estimated to be about seven orders of magnitude smaller than the CMB signal, and therefore unlikely to be directly detectable at present.
This is in line with other numerical calculations, and also with Planck observations of the CMB, which at the present accuracy do not exhibit a frequency-modulated signal from the recombination spectrum. Here this result has been obtained in a novel nonlinear diffusion model that is analytically solvable and offers a transparent description of the partial thermalization process. {The model is also expected to be useful regarding other astrophysical or cosmological equilibration processes in the non-thermal universe. An example is the reheating following inflation, which could be modelled using the nonlinear diffusion equation in case the timescale turned out to be finite.} 




\backmatter

\bmhead{Acknowledgments}

I thank Johannes H\"olck and Ralf Hofmann for a critical reading of the manuscript, for useful remarks, and John C. Mather for kindly transmitting the COBE-FIRAS data shown in Figure 6.

\section*{Declarations}

\begin{itemize}
\item Funding

The article has no external funding sources.
\item Conflict of interest/Competing interests

I have no competing interests. 
\item Ethics approval 

Not applicable
\item Consent to participate

Not applicable
\item Consent for publication

Not applicable
\item Availability of data and materials

The COBE-FIRAS data that are compared with my model calculations in Fig. 6 were provided by John C. Mather.
The model results shown in Figs. 1-6 (Mathematica-plots) are available as tables from the
corresponding author upon reasonable request.
\item Code availability 

The Mathematica-codes for the calculations shown in Figs. 1-6 can be made available upon request.

\item Authors' contributions

Single author GW of the manuscript. 
\end{itemize}
\end{document}